\documentclass[preprint,12pt]{elsarticle}
\usepackage[utf8]{inputenc}
\usepackage{graphicx}
\usepackage{url}
\usepackage{amsmath}
\usepackage{color}

\newcounter{bla}

\journal{Computer Physics Communications}

\begin{document}

\begin{frontmatter}


\title{Contact-based molecular dynamics of structured and disordered proteins
in a coarse-grained model: fixed contacts, switchable contacts and those
described by pseudo-improper-dihedral angles}

\author[a]{{\L}ukasz Mioduszewski\corref{author}}
\author[b]{Jakub Bednarz}
\author[b]{Mateusz Chwastyk}
\author[b]{\fbox{Marek Cieplak}}

\cortext[author]{Corresponding author. \textit{E-mail address:} l.mioduszewski@uksw.edu.pl}
\address[a]{Faculty of Mathematics and Natural Sciences, Cardinal Stefan Wyszyński University, Wóycickiego 1/3, 01-938 Warsaw, Poland}
\address[b]{Institute of Physics, Polish Academy of Sciences, Al. Lotników 32/46, 02-668 Warsaw, Poland}

\begin{abstract}
We present a coarse-grained C$_\alpha$-based protein model that can be 
used to simulate structured, intrinsically disordered and partially disordered proteins.
We use a Go-like potential for the structured parts and
two different variants of a transferable potential for the disordered parts.
The first variant uses dynamic structure-based (DSB) contacts that form and disappear 
quasi-adiabatically during the simulation. 
By using specific structural criteria we distinguish 
sidechain-sidechain, sidechain-backbone and backbone-backbone contacts.
The second variant is a non-radial multi-body pseudo-improper-dihedral (PID) potential
that does not include time-dependent terms but requires more computational resources. 
Our model can simulate in reasonable time thousands of residues on millisecond time scales.
\end{abstract}

\begin{keyword}
coarse grained model; intrinsically disordered proteins; molecular dynamics; implicit solvent; contact map;
\end{keyword}

\end{frontmatter}



{\bf PROGRAM SUMMARY}

\begin{small}
\noindent
{\em Program Title: cg}                                     \\
{\em Licensing provisions: MIT }            \\
{\em Programming language: C++ (new version), Fortran (old version)}                            \\
{\em Nature of problem(approx. 50-250 words):}\\

  Simulations of one or more protein chains, structured, intrinsically disordered or
  partially disordered. Calculating their equilibrium and kinetic properties in processes 
  including but not limited to: folding, aggregation, conformational changes, 
  formation of complexes, aggregation, response to deformation. 
  All those processes need long simulation times or many trajectories to properly sample the system.\\
\\
{\em Solution method(approx. 50-250 words):}\\

The simulations use a molecular dynamics implicit-solvent coarse-grained model where each pseudoatom represents 
one amino acid residue. Residues are harmonically connected to form a chain. 
The system evolves according to Langevin dynamics. The backbone stiffness potential 
involves bond angle and dihedral angle terms (or a chirality term in the structured case). 
Residues interact via modified Lennard-Jones or Debye-Hueckel potentials. 

The potential is different for structured and disordered parts of a protein.
A Go model contact map is used for the structured parts, where an interaction 
between two residues is attractive if effective spheres associated with
their heavy atoms overlap in the native structure [1,2,3,4]. 
Non-attractive contacts use only the repulsive part of the Lennard-Jones potential. 
The potential for the disordered parts has two versions:
in the quasi-adiabatic Dynamic Structure-Based (DSB) variant, the contacts are formed dynamically based on the structure of the chain 
and are quasi-adiabatically turned on and off [5,6]. In the slower, but more accurate Pseudo-Improper-Dihedral (PID)
variant the Lennard-Jones term is multiplied by a pseudo-improper-dihedral term that gives similar 
geometric restrictions with a time-independent Hamiltonian [7]. 

Proteins can be pulled by their termini to mimic stretching by an Atomic Force Microscope [8,9]. 
Multiple protein chains can be put in a simulation box with periodic boundary conditions 
or with walls that may be repulsive or attractive for the residues. The walls may move 
to mimic pulling or shearing deformations. The program accepts PDB files or protein 
sequences with optional user-defined contact maps. It saves the results in an output 
file (summary), a PDB file (structure) and a map file (contact map in a given snapshot).\\

{\em Additional comments including Restrictions and Unusual features (approx. 50-250 words):}\\
The program supports openMP (effective up to 8 cores, use 4 cores for optimal speed-up). The C++ version is hosted on Github, the Fortran version is distributed as a .tar archive.

\end{small}

\section{Usage}
There are two versions of the program: in C++ (new) and in FORTRAN (old). They should give the same results (up to numerical accuracy), with the exception of boundary conditions in the form of a simulation box with attractive walls (see section \ref{simulbox}). The C++ version is well documented online (\url{https://vitreusx.github.io/pas-cg/}). The rest of this paper refers to the old, Fortran version.

\begin{figure}[h!]
\centering
\includegraphics[width=\textwidth]{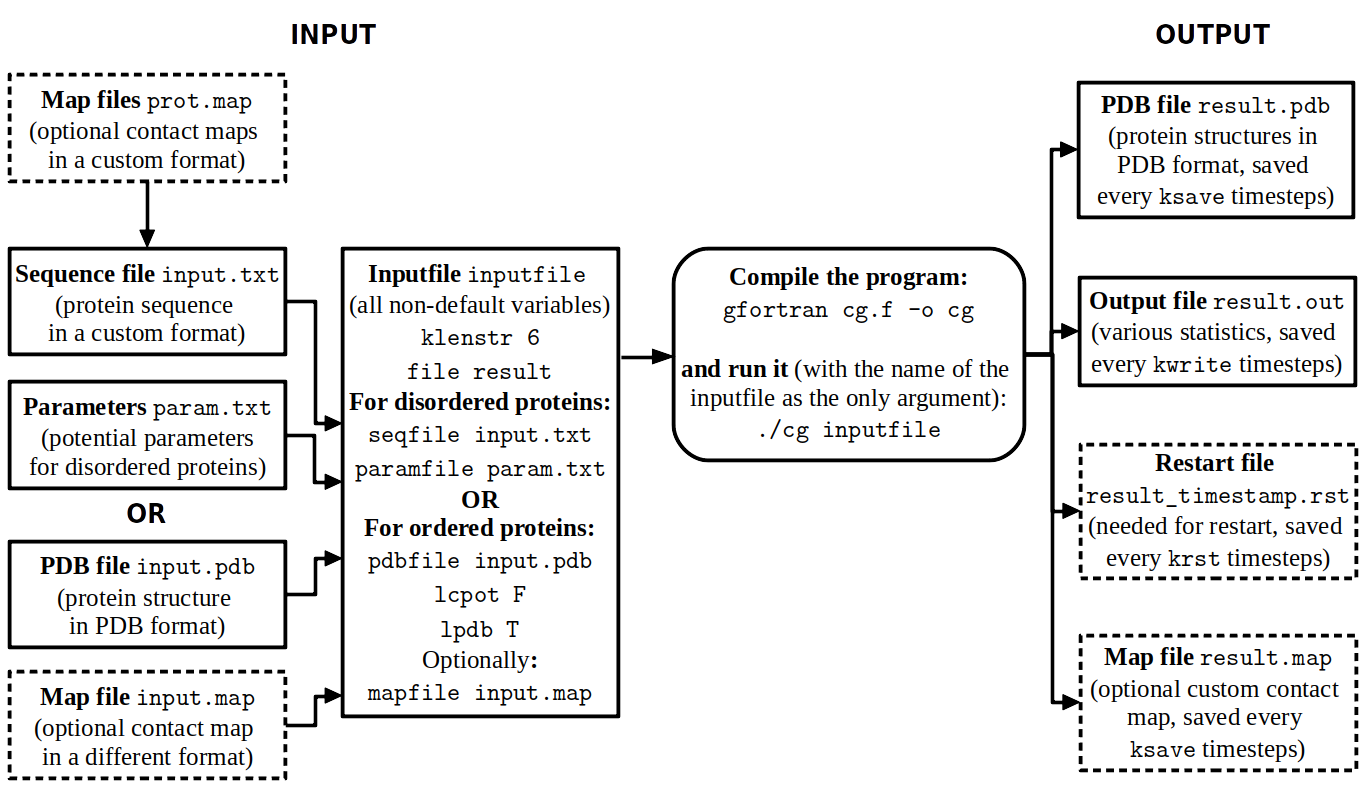}
\caption{A scheme visualizing the usage of the program.}
\label{scheme}
\end{figure}

The program is a single FORTRAN file (\texttt{cg.f}). After compilation, it requires exactly one argument: the name of the inputfile. Names of all the other files that may be needed must be written in the inputfile. Possible input and output files are shown in Fig. \ref{scheme}.
All files used in the program are in the plain text format. Each line of the inputfile has two words separated by a single space: the name of a variable and the value it should have. All variables have a predefined default value, so only the non-default variables need to be included in the inputfile. More technical details (like the list of the variables) are summarized in the \texttt{README.txt} file. This program is distributed as a \texttt{.tar} archive. 

\section{Equations of motion}
Each pseudoatom is centered at the position of the C$_\alpha$ atom of the amino acid it represents.
We use Langevin dynamics with the following equation of motion for the $i$-th residue in a chain:
\begin{equation}
m\frac{d^2\vec{r}_i}{d^2t}=\vec{F}_i-\gamma\frac{d\vec{r}_i}{dt}+\vec{\Gamma_i}
\end{equation}
where $m$ is the average amino acid mass 
(the variable \texttt{lmass} allows to turn on different amino acid masses), 
$\vec{r}_i$ is the position of the residue, 
$\vec{F}_i$ is the force resulting from the potential, 
$\gamma=2m/\tau$ is the damping coefficient and 
$\vec{\Gamma_i}$ is the thermal white noise with 
the variance $\sigma_T^2=2\gamma k_B T$. 
The implicit solvent is implemented via this noise and the damping.
The default damping coefficient $\gamma=2m/\tau$ is chosen to represent an 
overdamped case with diffusional (not ballistic) dynamics. 
More realistic
values of $\gamma$ should be about 25 times larger \cite{Hoang1,Veitshans},
but adopting them would lead to a much longer conformational dynamic.

The time unit $\tau \approx 1$ ns was verified for a different model 
with the same equations of motion \cite{szymczak}, so the value of $\tau$ 
is only orientational.
The equations of motion are integrated
by the fifth order predictor-corrector algorithm \cite{Tildesley,Haile}.
One $\tau$ corresponds to 200 integration steps.

The energy is expressed in units of $\epsilon$, 
where $\epsilon\approx 1.5$ kcal/mol \cite{Sikora,Poma2}.
The internal unit of energy in the code is also $\epsilon$. 
The distances in the input and output of the program are in the units of {\AA}.
However, the distances used in the integrating module are dimensionless: they
are divided by 5 {\AA}; hence the conversion unit $\texttt{unit}$ of 5.
(so a distance ,,10 {\AA}'' in the input 
is processed as ,,2'' during the integration of the equations of motion).

The residues in one chain are tethered harmonically with the spring constant 
$k=100$ {\AA}$^{-2}\cdot\epsilon$ and the equilibrium distance between them
is taken as 3.8 {\AA}.
If the simulation is based on a PDB file, 
equilibrium distances are taken from that file and can be different from 3.8 {\AA}.
Due to limitations in the chain numbering in the PDB files, the program
allows only 153 unique chain labels in a single file. 
However, there are ways to use more than one file \cite{capsid,virwolek}.
In Fortran, logical values true and false are denoted as T and F, respectively.
Setting the variable \texttt{lwritexyz} to T causes the program to use a much simpler XYZ format, which does not distinguish between chains (and prints only the x,y,z positions of each residue).

\section{Backbone stiffness}
Local interactions between amino acids separated by at most three residues in the protein sequence are known 
to stabilize folded proteins and improve their folding kinetics \cite{hoangstiffness}, and are even more 
important in determining the conformational ensembles of disordered proteins \cite{stelzl}. 
Those local interactions require using special backbone stiffness potentials (although in this program 
residues $i$ and $j \ge i+3$ can also attract each other in non-local interactions). 
Below we describe those potentials. Unstructured parts of a chain can use only bond angle and dihedral angle terms.

\subsection{Chirality potential for structured parts}
Chirality potential is defined \cite{wolek2} as 
$V^{CH}=\frac{1}{2}e_{chi}\Sigma_{i=2}^{i=N-2}(C_i-C^n_i)^2$ 
where $N$ is the number of residues and the chirality of the residue $i$ is 
$C_i=(\vec{v}_{i-1}\times\vec{v}_{i})\cdot\vec{v}_{i+1}/d_0^3$ 
where $\vec{v}_{i}=\vec{r}_{i+1}-\vec{r}_i$ and $d_0$ is equal to the length 
of $\vec{v}_{i}$ in the native structure (usually very close to 3.8~\AA). 
$C^n_i$ is the chirality of residue $i$ in the native structure. 
This dependence on the native structure restricts usage of the chirality potential only to structured proteins 
(the variable \texttt{lpdb} must be set to T). This potential can be turned on by setting the variable 
\texttt{lchiral} to T. The amplitude $e_{chi}$ of the potential can be tuned using the variable \texttt{echi} 
(the default value is 1, in the internal units of energy $\epsilon\approx$ 110 pN$\cdot$\AA).
It is computationally faster than the backbone stiffness potential based on the bond and dihedral angle terms 
defined below, but it is important to note that using chirality potential (with the default \texttt{echi}) lowers 
the effective room temperature of the system to 0.25 (in the program units, which correspond to $\epsilon/k_B$). 
The room temperature for structured proteins corresponds to the minimum of the median folding time and other 
properties of the system, defined in \cite{wolek2}.

\subsection{Bond and dihedral angle potentials}
The bond angle between residues $i-1,i,i+1$ is defined as $\theta_i$, and the dihedral angle between residues 
$i-2,i-1,i,i+1$ is $\phi_i$. Backbone stiffness potentials based on those angles are turned on by 
the variable \texttt{langle}.
\subsubsection{Structured parts}
The bond angle potential for the structured parts is harmonic with the minimum in the bond angle $\theta_n$ 
present in the native structure: 
$V_\theta=k_\theta(\theta-\theta_n)^2$, where the value of $k_\theta$ is 
defined by the variable \texttt{CBA} (its default value is 30, in $\epsilon/
$rad$^2$ units). 

The bond angle term is usually used together with the dihedral term, which 
can be turned on by setting the variable \texttt{ldih} T. The dihedral term 
has two forms. If the variable \texttt{ldisimp} is set to T, the harmonic 
form is used: $V_\phi^{simp}=\frac{1}{2}k_\phi(\phi-\phi_n)^2$, 
where $\phi_n$ is the dihedral 
angle in the native structure and $k_\phi$ can be set in the variable 
\texttt{CDH} (with the default value of 3.33, in $\epsilon/$rad$^2$ units).
The room temperature for this set of parameters has not been determined.

When \texttt{ldisimp} is F, a more sophisticated dihedral potential is used: 
$V_\phi= K_1[1-\cos(\phi-\phi_n)] + K_3[1-\cos(3(\phi-\phi_n))]$, where 
the $K_1$ and $K_3$ parameters are specified by variables \texttt{CDA} and \texttt{CDB}, 
both with a default value 0.66 (in $\epsilon/$rad$^2$ units), respectively. 
This set of values (together with the bond angle potential) corresponds to room temperature 0.7, at which the kinetics of folding of structured proteins is optimal \cite{wolek2}.

\subsubsection{Unstructured parts}
For the backbone stiffness of the unstructured parts we use a 3-letter alphabet 
distinguishing Gly, Pro and X residues, where X is one of the 18 other 
amino acids. For the bond angle we check only what the second and third 
residues are in the three consecutive residues forming the angle, which 
gives 9 possible choices. Including the first residue would give 27 
possibilities, but the symmetry between the first and the third residue in a 
bond angle can be broken (because proteins are chiral). The bond angle 
potential has the form of a sixth degree 
polynomial. Its coefficients for all 9 cases are listed in the parameter file 
specified by the variable \texttt{paramfile}.

The dihedral angle requires a consideration of four consecutive residues
but we distinguish only the two central ones, which again gives 9 possible combinations. 
The dihedral potential is governed by a formula
$a \sin( \phi )+b \cos( \phi )+c \sin^2( \phi )+d \cos^2( \phi )+e \sin( \phi ) \cos( \phi )$,
where the values of the coefficients $a,b,c,d,e$ for all 9 cases are also 
listed in the \texttt{paramfile}.

The coefficients for both the bond angle and the dihedral potentials 
were obtained \cite{lmc2018} by fitting the formulas to a potential resulting 
from the inverse Boltzmann method applied to a random coil database. 
This potential was shown to correctly describe properties of proteins
in denaturating conditions \cite{cglocalpot}. Thus our backbone stiffness 
has no preference towards any secondary structure, but $\alpha$-helices and $\beta$-sheets can still be 
formed by the backbone-backbone contacts (see section \ref{idps}).

Ghavami et al \cite{cglocalpot} distinguished only the middle residue and whether
it precedes proline, so only 6 from the 9 combinations in the \texttt{paramfile} 
are different.

The backbone stiffness is important in setting the effective room temperature for the model \cite{wolek2}, so 
in this case we checked what temperature corresponds best to the experimental end-to-end distances for 
the polyprolines of various lengths \cite{polyP}. We chose polyproline because its conformation is mostly 
determined by its backbone stiffness, so the parameters unrelated to the backbone stiffness could be optimized 
in the correct temperature. Temperatures between 0.35 and 0.4 $\epsilon/k_B$ turned out to be optimal, which is 
consistent with our energy unit $\epsilon\approx 1.5$ kcal/mol. Then 0.38 $\epsilon/k_B \approx 300$ K.

\subsubsection{User-defined potential}
The backbone stiffness potential may be also specified in a table where each row corresponds to a different angle and 
each column to a different Gly/Pro/X combination (in the same manner as in the previous section). The file 
\texttt{stiffnesstabularized.txt} contains the Boltzmann inversion potentials from Ghavami et al \cite{cglocalpot}.

\section{Non-local potential}
\begin{figure}[h!]
\centering
\includegraphics[width=\textwidth]{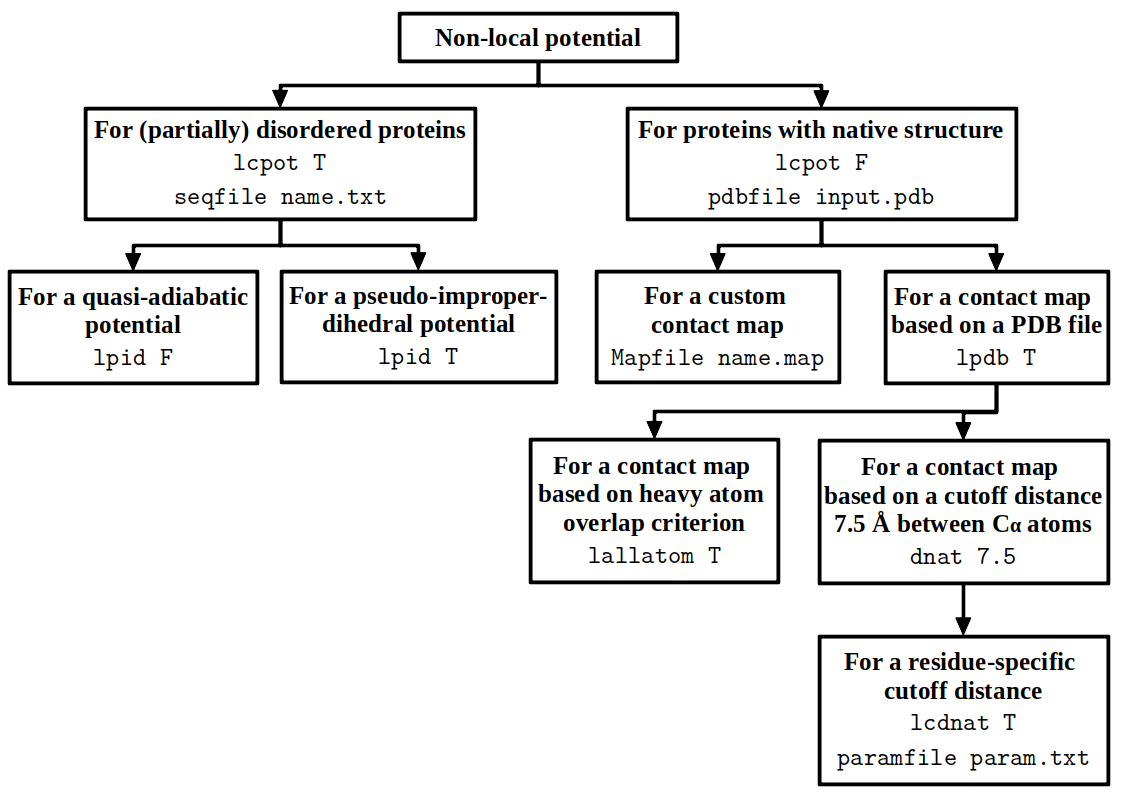}
\caption{A scheme showing variants of the non-local potential.} 
\label{potentials}
\end{figure}

The non-local potential is different for structured and disordered proteins. Possible variants of the potential together with the variables needed to switch between them are schematically shown in Fig. \ref{potentials}.
\subsection{Structured parts}
If the variables \texttt{lpdb} and \texttt{lallatom} are set to T and a PDB file is provided, the program 
automatically constructs from the PDB file a contact map based on the overlap of heavy atoms \cite{Settani} after 
assigning to each of them a radius that takes into account different positions of atoms in the amino acid (e. g. it 
distinguishes the radii of C$_\alpha$ and C$_\beta$ atoms) \cite{Tsai}. In the case of an incomplete PDB file a simpler 
contact map based on C$_\alpha$-C$_\alpha$ distances may be used (by setting \texttt{lallatom} F).

An analysis of 62 variants of the Go-like models has identified four that
show the best agreement with the experimental stretching of structured
proteins by means of tips mounted in the Atomic Force Microscopes (AFM).
Among the four there is a simple variant in which each attractive contact
is described by a Lennard-Jones (L-J) potential with the depth 1 $\epsilon$ \cite{survey}.

It is important to note that the overlap of heavy atoms is just one possible criterion for defining a contact. 
More sophisticated criteria may involve the chemical identity of the overlapping heavy atoms \cite{rcsu}. 
Constructing such contact maps is not possible in the program presented here, but we provide a webserver for 
computing contact maps using a rCSU method: \url{info.ifpan.edu.pl/~rcsu/}

The simple L-J potential describing an attractive interaction between residues $i$ and $j$ has the form 
$\epsilon\left[{\left(\frac{r_{min^{ij}}}{r}\right)}^{12}-2\left(\frac{r_{min^{ij}}}{r}\right)^6 \right]=
4\epsilon \left[{\left(\frac{\sigma_{ij}}{r}\right)}^{12}-\left(\frac{\sigma_{ij}}{r}\right)^6 \right]$, 
where $r_{min^{ij}}=2^{1/6}\sigma_{ij}$ is the minimum of the potential (for the structured proteins it is 
equal to the C$_\alpha$-C$_\alpha$ distance in the native conformation). The residues that are not in a 
contact repel each other with a truncated and shifted form of the L-J potential:
$V_r(r \leq r_{o})=\epsilon\left[{\left(\frac{r_{min}}{r}\right)}^{12}-2\left(\frac{r_{min}}{r}\right)^6 + 1\right]$, 
where $r_{o}=4$ \AA~ ensures the excluded volume (by default it is set to 5 \AA~ in the variable \texttt{cut}). 
Contacts between the $i$th and $i+2$nd residues are always described by the repulsive version of the L-J potential 
(also in potentials for disordered proteins). Please set \texttt{cut} to 4 {\AA} when using only the 
structure-based potential, in order to get results consistent with the literature \cite{Sikora,wolek2,survey}.

\subsection{Unstructured parts}
\label{idps}
\begin{figure}[h!]
\centering
\includegraphics[width=\textwidth]{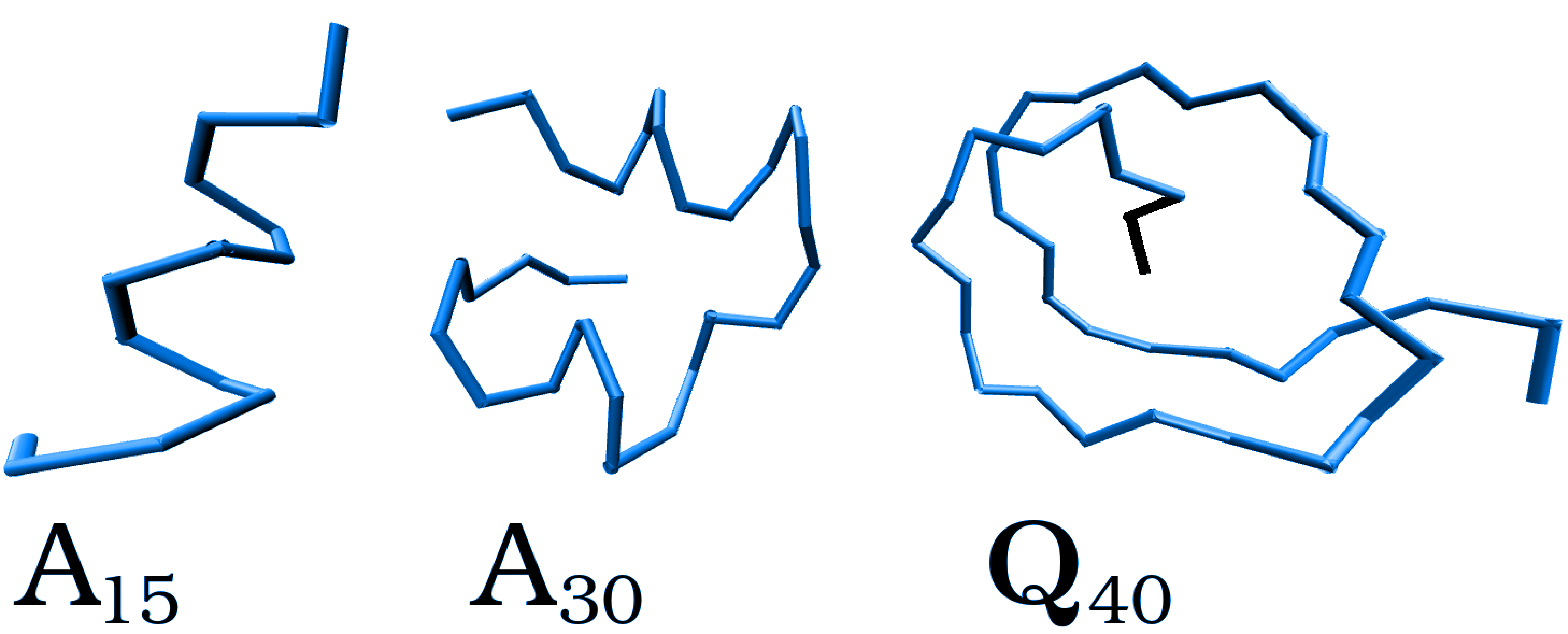}
\caption{Three examples of conformations generated by the quasi-adiabatic potential \cite{lmc2018}. }
\label{secondary}
\end{figure}

The backbone stiffness potential for disordered parts does not favor any 
secondary structure, so in order to make structures like $\alpha$-helices 
or $\beta$-sheets possible, we allow for attractive contacts between the $i$th and $i+3$rd residues in the chains, 
as these contacts correspond to hydrogen bonds between backbone atoms in the all-atom representation \cite{kolinski}. Forming backbone-backbone contacts results in helix-like and sheet-like structures that are similar to $\alpha$-helices and $\beta$-sheets (see Fig. \ref{secondary}). In this approximate way, our model incorporates secondary structure elements that may be disrupted and reformed, as backbone-backbone contacts keep being broken and reestablished.
However the nature of $i,i+4$ contacts is different: the quasi-adiabatic 
potential works best with $i,i+4$ contacts turned on (the variable \texttt{lii4} set to T),
and the pseudo-improper-dihedral potential without them (\texttt{lii4} F). A more detailed comparison of the model variants is available in \cite{lmc2020}.

The logical variable \texttt{lcpot} controls if our custom potential for disordered proteins is turned on. With 
\texttt{lcpot} F only residue pairs listed in the contact map attract each other. Setting \texttt{lcpot} T allows 
for a custom potential for all the residues, except those present in user-defined contact maps. So if you provide a PDB 
file and tell the program to construct a Go model contact map for it (by specifying \texttt{lpdb} T), but at the same 
time you set \texttt{lcpot} T, the pairs of residues listed in the contact map will interact via the ,,simple'' L-J 
potential described above, but those not in the contact map will be subject to a potential customized for intrinsically 
disordered proteins. Therefore, the \texttt{lcpot} option should also be used for proteins that are only partially disordered. There are two versions of the disordered potential: quasi-adiabatic and pseudo-improper-dihedral. Both of them 
are described below.

\subsubsection{Quasi-adiabatic potential}
\label{quasi}
The default, quasi-adiabatic version is faster, and uses contacts described by the L-J potential with 
the depth 1 $\epsilon$ (just as the Go potential for the structured parts) and a minimum $r_{min}$ (defined below), 
but here the contacts are formed and broken quasi-dynamically, based on the geometry of the chain. 
As real amino acids are made from a sidechain and a backbone, in our model we distinguish sidechain-sidechain (ss), 
backbone-sidechain (bs) and backbone-backbone (bb) contacts (each has different criteria for formation, but only 
one type of contact can be made between residues at a given time).

After conditions for forming the contact are met, formation of the contact happens quasi-adiabatically: the depth of 
the L-J potential well increases linearly form 0 to $\epsilon$ within 10$\;\tau$, which is long enough for the system 
to thermalize, but short enough to not disrupt dynamics of the system \cite{lmc2018}. Breaking the contact also 
happens quasi-adiabatically, with the same time scale.

There are 3 criteria for forming a contact between residues $i$ and $j$:
\begin{itemize}
\item \textbf{Distance}: each type of contact has specific $r_{min}$ associated with it (based on a maximum of a 
statistical distribution from the PDB \cite{lmc2018}). It is 5 \AA~ for bb, 6.8 \AA~ for bs, and a value specific for a 
given pair of amino acids for an ss contact. Those numbers are specified in the \texttt{paramfile}. To make a contact of 
a given type, the distance $r_{i,j}$ between the residues must be smaller than $r_{min}\cdot (1+tolerance)$  to make a 
contact (the default value of the variable \texttt{tolerance} is 0). 
\item \textbf{Direction}: the approximate directions of a backbone hydrogen bond or a sidechain C$_\beta$ atom may be 
derived based on the position of 3 consecutive C$_\alpha$ atoms \cite{rey,hoang,lmc2018}. Let's assume that $v_i$ is the 
vector connecting the $i$th and $i+1$st beads (centered on the supposed positions of the C$_\alpha$ atoms). Then we can 
construct two auxiliary vectors:
\begin{align}
{\bf n}_i = \frac{{\bf v}_{i} - {\bf v}_{i-1}}{{\bf v}_{i} - {\bf v}_{i-1}}
&&
{\bf h}_i=\frac{{\bf v}_{i} \times {\bf v}_{i-1}}{|{\bf v}_{i} \times {\bf v}_{i-1}|} 
\end{align}
They represent the normal (${\bf n}_i$) and binormal (${\bf h}$) vectors of the Frenet frame defined by the 
${\bf v}_{i}, {\bf v}_{i-1}$ vectors, and can be associated with the direction of the sidechain ($-{\bf n}_i$) 
and with the direction of a hydrogen bond made by an atom from the backbone ($\pm {\bf h}_i$). A survey of the 
contacts from the PDB \cite{rey,hoang} led to the following criteria for a contact of each type:

\begin{itemize}
\item bb: $|cos({\bf {h}}_i,{\bf {r}}_{i,j})| > 0.92$, $|cos({\bf {h}}_j,{\bf{r}}_{i,j})| > 0.92$ 
(the threshold angle is 23$^o$), $|cos({\bf {h}}_i,{\bf {h}}_j)| > 0.75$ (the treshold angle is 41$^o$)
\item bs: $\cos({\bf n}_i,{\bf r}_{i,j})<0.5$, 
 $|\cos({\bf h}_j,{\bf r}_{j,i})|>0.92$ 
 (where the $i$th residue donates its sidechain and the $j$th residue its backbone)
\item ss: $\cos({\bf n}_i,{\bf r}_{i,j})<0.5$, $\cos({\bf n}_j,{\bf r}_{j,i})<0.5$ (the threshold angle is $60^o$)
\end{itemize}
\item \textbf{Coordination number}: the solvent in the program is implicit, so we cannot simulate how polar residues 
make contacts with water molecules. Using the one-bead-per-residue model also makes the protein less dense than it 
really is. We try to atone for that by allowing each amino acid to form only a limited number of contacts. Each 
residue can make only 2 contacts with their backbone (1 for proline) and $s$ ss contacts with their sidechain. 
Out of these contacts only $n_H$ can be made with hydrophobic residues and $n_P$ with polar residues. 
$s$, $n_H$ and $n_P$ depend on the amino acid. Those numbers and the division into polar and hydrophobic residues 
are available in the \texttt{paramfile} and here in the table \ref{aaproperties}. Those values are based on the PDB 
statistics \cite{lmc2018}. bs contacts are considered to be ,,polar'' (thus $n_H+n_P$ may be bigger than $s$).

\begin{table}[h!]
\begin{center}
\begin{tabular}{|l|llllllllll|}
\hline
name & {\bf Gly} & {\bf Pro} & {\bf Gln} & {\bf Cys} & {\bf Ala} & {\bf Ser} & {\bf Val} & {\bf Thr} & {\bf Ile} & {\bf Leu}\\
type & - & - & P & P & H & P & H & P & H & H\\
$s$   & 0 & 0 & 2 & 3 & 3 & 2 & 4 & 2 & 5 & 5\\
$n_H$ & 0 & 0 & 0 & 2 & 1 & 0 & 4 & 0 & 4 & 4\\
$n_P$ & 0 & 0 & 2 & 2 & 1 & 2 & 1 & 2 & 2 & 2\\ \hline 
name & {\bf Asn} & {\bf Asp} & {\bf Lys} & {\bf Glu} & {\bf Met} & {\bf His} & {\bf Phe} & {\bf Arg} & {\bf Tyr} & {\bf Trp}\\
type & P & P$_-$ & P$_+$ & P$_-$ & H & P & H & P$_+$ & H & H\\
$s$   & 2 & 2 & 2 & 2 & 4 & 2 & 6 & 2 & 4 & 5\\
$n_H$ & 0 & 0 & 0 & 0 & 1 & 0 & 4 & 0 & 2 & 4\\
$n_P$ & 2 & 2 & 2 & 2 & 1 & 2 & 2 & 2 & 2 & 3\\ \hline
\end{tabular}
\caption{Coordination numbers of residues. ,,type'' indicates whether a given amino acid is considered polar (P) 
or hydrophobic (H). The subscripts + or -- indicate the charge.
$s$, $n_P$ and $n_H$ are defined in the text.}
\label{aaproperties}
\end{center}
\end{table}
\end{itemize}

\begin{figure}[h!]
 \centering
 \includegraphics[width=0.72\textwidth]{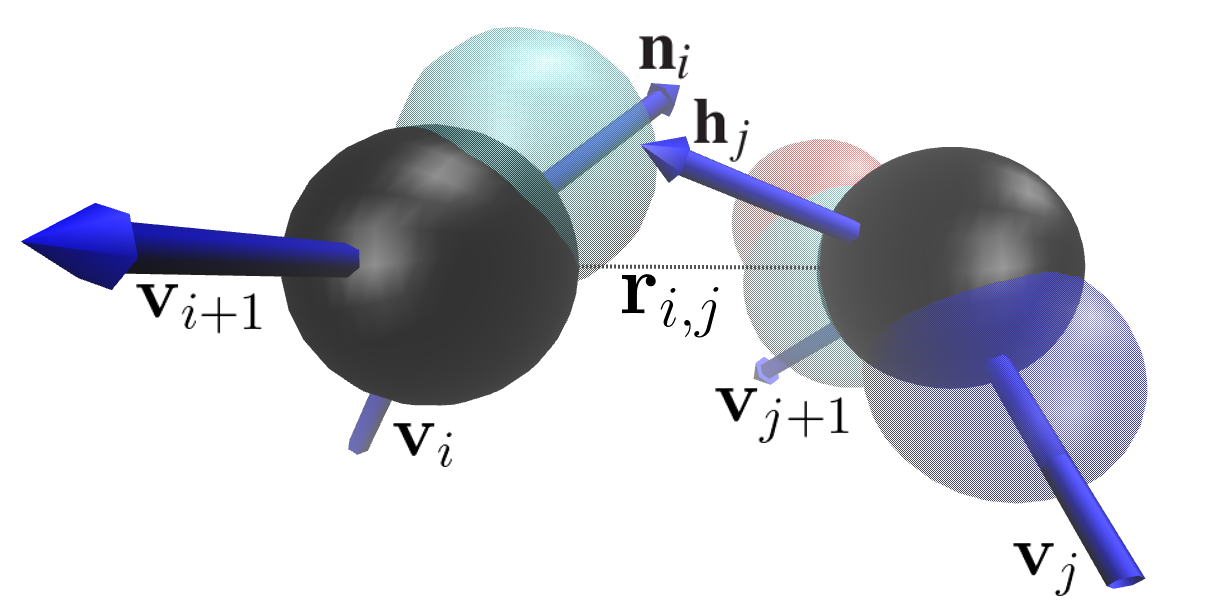}
 \caption{An example of a bs contact between Ala (on the left) and Met.
Only sidechain heavy atoms of Ala and only backbone heavy atoms of Met are shown, all in the CPK coloring scheme. 
The C$_\alpha$ atoms are shown in black.
The arrows indicate vectors ${\bf {v}}$, ${\bf {h}}$ and $-{\bf {n}}$. The vector $r_{i,j}$ connecting the 
C$_\alpha$ atoms is drawn as a dotted line.
The value of $r_{min}$ for bs contacts is 6.8 {\AA}.} 
 \label{arrows}
 \end{figure}

An example of a bs contact is shown in Fig. \ref{arrows}.
 
The contact between residues $i$ and $j$ is considered broken when 
$r_{i,j} > \texttt{cntfct} \cdot 2^{-1/6}\cdot r_{min}$, where the default value of the variable \texttt{cntfct} is 
and the 2$^{-1/6}$ factor comes from the inflection point of the L-J potential. 

Notice that despite complex geometrical criteria required for forming a contact in the quasi-adiabatic potential, 
the contact itself is described by a spherically-symmetric potential. 

It is important to note that the \texttt{cntfct} variable is also used in another context, to 
describe the breaking of a contact in the structure-based Go model. In this case the condition 
$r_{i,j} > \texttt{cntfct} \cdot 2^{-1/6}\cdot r_{min}$ does not change anything in the dynamics of the model. 
It is just used to count how many contacts are present. Using this criterion in this way is arbitrary, 
but we recommend using \texttt{cntfct=1.5} for counting contacts in the Go model, to get the same results as 
in the literature \cite{Sikora,wolek2,survey}.

\subsubsection{Pseudo-improper-dihedral potential}
\label{pidmodel}
The second custom potential (turned on by setting the variable \texttt{lpid} T) is slower, but its hamiltionian is 
time-independent and it better agrees with experiment \cite{lmc2020}. It distinguishes only backbone-backbone (bb) and 
sidechain-sidechain (ss) contacts. A contact between residues $i$ and $j$ is described by a 6-body potential that 
involves residues $i-1,i,i+1$ and $j-1,j,j+1$. It is a product of a Lennard-Jones potential and two imroper dihedral 
potentials, each of them involving only 4 residues. A dihedral angle potential always restricts the angle between two 
planes, each set by three residues.
Such a potential is already used in the model to describe the backbone stiffness 
(then it acts on 4 consecutive residues).
In its ,,improper'' version, this potential acts on three consecutive beads and a sidechain bead connected to the 
middle one. The ,,pseudo-improper'' version acts on residues that are not even covalently connected. When residues 
$i$ and $j$ interact, the first ,,pseudo-improper-dihedral'' (PID) angle is defined by the planes spanned by 
residues $i-1,i,i+1$ and $i-1,j,i+1$. The second PID angle is between the planes $j-1,j,j+1$ and $j-1,i,j+1$. 

\begin{figure}[h!]
\centering
\vspace{-10pt}
\includegraphics[width=0.9\textwidth]{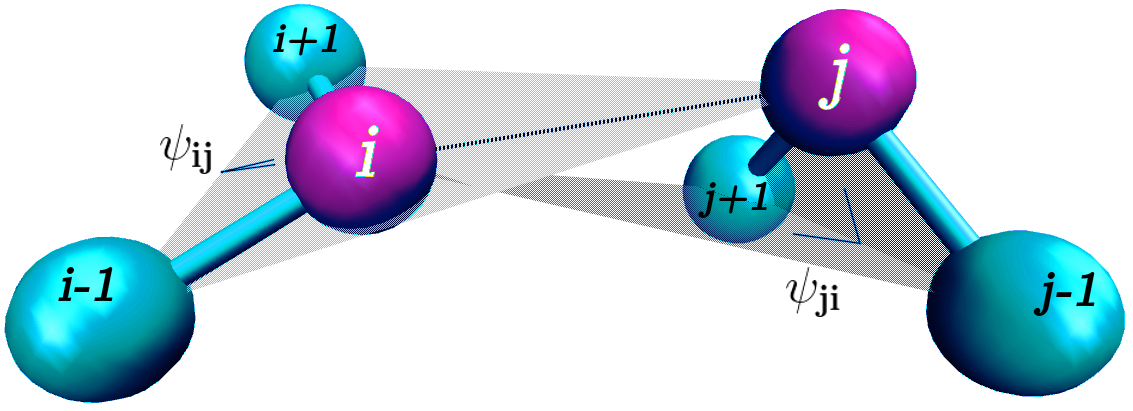}
\vspace{-10pt}
\caption{The idea of the PID angles. The interaction between residues $i$ and $j$ involves angles $\Psi_{ij}$ 
(defined by $i-1,i,i+1$ and $j$ beads) and $\Psi_{ji}$ (defined by $j-1,j,j+1$ and $i$ beads).}
\label{idea}
\end{figure}

Those planes are shown in Fig. \ref{idea}. The potential for each pair of residues is described by one distance 
$r_{i,j}$ and two pseudo-improper-dihedral (PID) angles 
$\Psi_{i,j}$ and $\Psi_{j,i}$. The potential is a product of three terms:
$V(\psi_{i,j},\psi_{j,i},r_{i,j})=\lambda_{i,j}(\psi_{i,j})\lambda_{j,i}(\psi_{j,i})\phi(r_{i,j})$, 
where $\phi(r_{i,j})$ is a L-J potential.
The PID angle distributions show clear peaks for the bb and ss cases. We used the Boltzmann inversion to discover 
that $\lambda(\psi)$ is indeed a Gaussian-like function \cite{lmc2020}. Because the peaks are narrow, instead of a 
Gaussian function we implemented two functions that vanish away from the peak (i.e. they have a compact support). The 
first of them is a fragment of a cosine function (used if the variable \texttt{lcospid} is T), shifted to the peak 
value $\psi_0$ and rescaled by the $\alpha$ parameter that is inversely proportional to the width of the peak:
\begin{equation}
\lambda(\psi) = \begin{cases}
0.5\cdot\cos{[\alpha(\psi-\psi_0)]}+0.5 &\text{when $-\pi < \alpha(\psi-\psi_0) < \pi$}\\
0 &\text{otherwise}
\end{cases}
\end{equation}

The second equation gives a very similar shape of $\lambda(\psi)$, but is made from algebraic sigmoid functions, 
so it is faster (here $\upsilon=\alpha(\psi-\psi_0)$):
\begin{equation}
0.5\cdot\cos{\upsilon}+0.5\approx
\frac{(\upsilon/\pi)^2-2\cdot |\upsilon/\pi|+1}{2\cdot (\upsilon/\pi)^2-2\cdot |\upsilon/\pi|+1}
\end{equation}

The distributions of the PID angles for the bb and ss cases are different so $V=V^{ss}+V^{bb}$.
For the ss contacts, $V^{ss}(\psi_A,\psi_B,r) =
\lambda^{ss}(\psi_A)\lambda^{ss}(\psi_B)\phi^{ss}(r)$. 
It turns out that the supports of $\lambda^{bb}$ and $\lambda^{ss}$ do not overlap. 
This makes the program faster (it computes only one $\lambda$ at a time). 

The value of $r_{min}$ for the ss cases depends on which pair of amino acids is interacting. 
All values of $r_{min}$ are listed in the \texttt{paramfile} (in the same place as for the quasi-adiabatic potential).

The bb PID angle distribution has two peaks ($\psi_0^{bb+}$ and $\psi_0^{bb-}$), corresponding to the anti-parallel 
and parallel $\beta$ strands, so the bb potential has two corresponding terms:
\begin{equation}
\lambda_{bb}(\psi) = \begin{cases}
0.5\cdot\cos{[\alpha^{bb+}(\psi-\psi_0^{bb+})]}+0.5 &\text{when $-\pi < \alpha^{bb+}(\psi-\psi_0^{bb+}) < \pi$}\\
0.5\cdot\cos{[\alpha^{bb-}(\psi-\psi_0^{bb-})]}+0.5 &\text{when $-\pi < \alpha^{bb-}(\psi-\psi_0^{bb-}) < \pi$}\\
0 &\text{otherwise}
\end{cases}
\end{equation}

These two terms correspond to a left- and right- handed $\alpha$-helix in the case of $i,i+3$ interactions, 
so only one of them is used to mimic right-handedness of most of the $\alpha$-helices in proteins.

The repulsive part of the LJ potential should always be 
present for small distances, to prevent the residues from passing through one another (the excluded volume effect). 
So even if $\lambda$ is set to 0, the repulsive potential should not vanish. Therefore:
\begin{equation}
V^{bb}(\psi_A,\psi_B,r) = \begin{cases}
\lambda^{bb}(\psi_A)\lambda^{bb}(\psi_B)\phi^{bb}(r) &\text{when $r > r_{min}^{bb}$}\\
\phi^{bb}(r)+(1-\lambda^{bb}(\psi_A)\lambda^{bb}(\psi_B))\epsilon^{bb} &\text{otherwise}
\end{cases}
\end{equation}

For the bb case, $r_{min}$ is stored in the variables \texttt{rbb1} and \texttt{rbb2}(corresponding to the 
$^+$ and $^-$ cases). The default values are $r_{min}^+=5.6$ \AA~ and $r_{min}^-=6.2$ \AA. Other default parameters are: 
$\alpha^{bb+}=6.4$, $\alpha^{bb-}=6.0$, $\alpha^{ss}=1.2$, $\psi_0^{ss}= -0.23$ rad, $\psi_0^{bb+} = 1.05$ rad, 
$\psi_0^{bb-} = -1.44$ rad. The fit to the statistical potential is available in ref. \cite{lmc2020}.

The depth of the L-J potential $\phi$ does not have to be uniform and always equal to 1 $\epsilon$. For bb contacts, 
$\epsilon^{bb}$ can be set in the variable \texttt{epsbb}. By default $\epsilon^{bb}=1$ $\epsilon$.
If the variable \texttt{lmj} is set to T, the values of $\epsilon^{ss}$ for each possible pair of amino acids are taken from a Miyazawa-Jernigan matrix \cite{MJ96}, which can be supplied by the 
\texttt{paramfile}. Otherwise $\epsilon^{ss}=1$ $\epsilon$. We include the \texttt{paramfile}s with the MDCG matrix derived from all-atom simulations 
\cite{mdcg} multiplied by 0.05 (\texttt{parametersMD05.txt}). 
The best agreement with the experiment for the PID model is accomplished for the MDCG matrix multiplied by a factor
between 0 and 0.1 \cite{lmc2020}. IDPs are strongly hydrated, which may be the cause of such small values. 
We include multiplier 0.05, because 0.1 turned out to lead to excessive aggregation of many chains 
(in ref. \cite{lmc2020}, the PID model is parameterized only for single chains).
In example 2 (attached with the program) we use $\alpha$-synuclein, a strongly hydrated IDP that was extensively studied \cite{robustelli,mcasyn}. The parameters used in example 2 should be a good starting point for recreating its properties (like the number of contacts), but further parameterization is probably needed.

Long sidechains have significant flexibility, resulting in broad distributions of C$_\alpha$-C$_\alpha$ distances 
for ss contacts \cite{lmc2018}. 
To recreate this we introduced a modified form of the L-J potential, shown below (it is turned on by the 
\texttt{lsink} variable, parameter $r_{min}^{bb}$ is specified by the variable \texttt{cut}, parameters $r_{min}^{ss}$ are different for each possible pair of amino acids and are taken from the \texttt{paramfile}):

\begin{equation}
\phi^{ss}(r)=\begin{cases}
\epsilon^{ss}\left[{\left(\frac{r^{ss}_{min}}{r}\right)}^{12}-
2\left(\frac{r^{ss}_{min}}{r}\right)^6 \right] &\text{when $r > r_{min}^{ss}$}
\\
-\epsilon^{ss} &\text{when $r_{min}^{ss} > r > r_{min}^{bb}$}
\\
\epsilon^{ss}\left[{\left(\frac{r^{bb}_{min}}{r}\right)}^{12}-
2\left(\frac{r^{bb}_{min}}{r}\right)^6 \right] &\text{when $r_{min}^{bb} > r$}
\end{cases}
\end{equation}

\subsubsection{Electrostatics}

The electrostatic interactions for the quasi-adiabatic and pseudo-improper-dihedral potentials are the same 
(unless the variable \texttt{lepid} is set to T, which is an untested feature). 
They are described by the Debye-Hueckel (D-H) screened electrostatic potential 
$V_{D-H}(r_{i,j})=\frac{q_iq_j\exp{(-r_{i,j}/s)}}{4\pi \varepsilon \varepsilon_0 r_{i,j}}$, 
where the screening length $s$ is defined in the variable \texttt{screend} and the charges $q_i, q_j$ are 
0, +1 or -1, based on the charge defined in the \texttt{paramfile} (in our units the elementary charge $e=1$). 

The electric permittivity $\varepsilon$ is constant if the variable \texttt{lecperm} is T, but if \texttt{lecperm} 
is F, a distance-dependent  $\varepsilon=4$ \AA$/r_{i,j}$ is used \cite{flap}. In both cases $\varepsilon$ 
(or the coefficient before $r_{i,j}^{-1}$) is included in the amplitude of the D-H interactions, \texttt{coul}. 
Default values of this amplitude are 85 $\epsilon \cdot$\AA$\cdot$\AA~ (for the distance-dependent 
$\varepsilon=4$ \AA$/r_{i,j}$) and 2.63 $\epsilon \cdot$\AA~ (for the constant $\varepsilon=80$ case). 

The electrostatic interactions between the oppositely-charged residues can be treated differently by setting the 
variable \texttt{lrepcoul} T. Then the attractive interactions between those residues are governed by the custom 
potential (quasi-adiabatic or pseudo-improper-dihedral). The repulsive interactions between same-charged residues 
are still treated by the D-H potential.

\subsection{Disulfide bonds}
For structured proteins, cysteines are treated as every other amino acid, with the exception of those specified in 
the SSBOND records in the PDB file - those are connected harmonically with the spring constant \texttt{H1} and the 
equilibrium distance 6 \AA. If the variable \texttt{lsslj} is T, even the cysteines in the SSBOND records are treated 
as other amino acids.

The options for the disordered case (when \texttt{lcpot} is T) work only for the quasi-adiabatic version of the model.
The default option is the same harmonic potential as above, but it is quasi-adiabatically turned on when the criteria of 
an ss contact formation are fulfilled, together with two additional conditions: cysteines are not part of any other 
disulfide bond, and the sum of the numbers of their neighbours must be smaller than \texttt{neimaxdisul} (9 by default). 
The number of neighbours is computed as the number of residues within a radius \texttt{rnei} (default 7.5 \AA) from a 
given cysteine. This harmonic potential can be quasi-adiabatically turned off if $(r_{ij}-6\text{ \AA})^2>0.1$ \AA$^2$ 
and the total number of neighbours is smaller than \texttt{neimaxdisul}. Thus, the disulfide bond formation in disordered 
proteins can be turned off by specifying \texttt{neimaxdisul}$=-1$. Those neighbour calculations are supposed to mimic 
the effect of solvent accessibility, which is crucial for the disulfide bond oxidation and reduction \cite{thirumalai}.

If the variable \texttt{lsselj} is T, a special L-J potential with the depth \texttt{dislj} (4 $\epsilon$ by default) 
is used. When this special type of contact is quas-adiabatically turned on between a pair of cysteines, they can no 
longer make such pairs with other cysteines. This contact obeys the same criteria for formation and breaking as any 
other ss contact.

\section{Simulation protocols, geometries and boundary conditions}

All the variables that describe time in the sections below are given in $\tau$ units. Every type of simulation can 
be repeated multiple times with different random seed during one execution of the program. Each repeat is called a 
trajectory. The variable \texttt{ntraj} determines the number of repeats. Out of all residues, the first \texttt{nen1} 
of them can be ,,frozen'' - they will not move during the simulation (which may greatly reduce computation time).

\subsection{Simulation protocols}

\subsubsection{Equilibrium simulations}
With the default options the proteins are put in an infinite space and there are no external forces acting upon them 
(except for friction and the thermostat). Time of the whole simulation is \texttt{mstep}. All the other protocols are 
preceded by the equilibration period. Boundary conditions may then be different, but the duration of the equilibration 
is always specified by the variable \texttt{kteql}. However the simulation time never exceeds \texttt{mstep}.

\subsubsection{Stretching by the tips of the AFM}
If the \texttt{lvelo} variable is T, the first and last residues are caught by harmonic springs with the spring 
constant \texttt{HH1} and pulled apart in the direction determined by the vector connecting both residues. 
To perform stretching simulations with spring constant resembling real AFM tips, please set \texttt{HH1} to 0.06 $\epsilon/$\AA$^2$ in the inputfile \cite{Sikora}.

The springs start to act after the equilibration and this vector is also determined at the time $t=$\texttt{kteql}. 
Then, the springs are moved apart from each other along this vector with the velocity \texttt{velo} 
(in \AA/$\tau$ units). The variable \texttt{lforce} has a similar effect, but the pulling occurs with a constant 
force (specified in the variable \texttt{coef}, in $\epsilon/$\AA~ units) acting on the terminal residues, 
instead of the constant speed.

\subsubsection{Folding simulations}

When the variables \texttt{lconftm} or \texttt{lmedian} are T, the simulation starts from a straight line or a 
self-avoiding random walk (which is controlled by the variable \texttt{lsawconftm}). Then, for \texttt{lconftm}, 
the average time of contact formation is computed for each native contact. The average is taken over all trajectories.

For \texttt{lmedian}, the program computes the median folding time. The folding time is defined as the time where all 
the native contacts are established \cite{wolek2,Zhao1}. If more than half of the trajectories did not end in the folded 
state, the simulation ends (because then it is impossible to compute the median folding time).
Trajectories during folding simulations may end after \texttt{mstep} timesteps, or after the native conformation 
is reached (if the variable \texttt{lnatend} is set to T).

\subsubsection{Unfolding simulations}

When the variable \texttt{lunfold} is T, the simulation starts from the native structure and the average time of 
contact breaking is computed for each native contact. The average is taken over all trajectories. Those simulations 
are usually performed in the temperature \texttt{temp} higher than the room temperature.
If the variable \texttt{lthermo} is T, thermodynamic properties such as the heat capacity 
$C_V=\frac{<E^2>-<E>^2}{T^2}$ are computed. A set of different temperatures can be simulated by specifying different 
\texttt{tstart} and \texttt{tend}.

\subsubsection{Deformations of the simulation box}
\label{simulbox}
When the variable \texttt{lwall} is T, the simulation does not occur in infinite space but in a box. 
Currently all the simulation protocols supported in this case require a specified density (in residue/\AA$^3$ units), 
so the variable \texttt{ldens} must be set to T. In the case of proteins from a PDB file, the box sizes in the 
CRYST1 record are initially used. For disordered proteins, the initial density is \texttt{sdens} and the simulation 
box is a cube. The chains (straight lines or self-avoiding walks, as set in \texttt{lsawconftm}) are constructed 
randomly in the box. If a chain goes beyond the box, the box size is increased to fit it (unless there are periodic 
boundary conditions in the initial state, which is controlled by the \texttt{lcpb} variable). It means that the actual 
density during the start of the dynamics may be even lower than \texttt{sdens}. Then, the box shrinks uniformly in all 
the 3 dimensions (the wall speed is \texttt{densvelo}) until it reaches the target density \texttt{tdens}. Then the 
box size does not change for a time equal to \texttt{ktrest}.

If the variable \texttt{loscillate} is T, the box starts to deform again (after the resting period specified in 
\texttt{ktrest}). There are two possible modes of oscillating the box walls: pulling and shearing (determined by the 
\texttt{lshear} variable). In the pulling mode, the distance between the two walls perpendicular to the Z axis changes 
periodically. In the shearing mode, the same two walls move periodically along the X axis.

The oscillation starts by achieving the maximal amplitude of oscillation. If the variable \texttt{lampstrict} is T, 
the amplitude \texttt{ampstrict} is the wall displacement (in the Z direction for pulling, in the X direction for 
shearing) in \AA. Otherwise, the amplitude \texttt{amp} is determined relative to the initial distance between the two 
moving walls.
After the walls are moved to their maximum displacement, there is another resting period (with a duration time 
\texttt{ktrest}) for equilibration. Then the displacement $s$ changes periodically:  
$s(t)=s_0(1-A\cos{2\pi (t/period)})$, where the period of the oscillations is the variable \texttt{period} and the 
amplitude $A$ is either \texttt{amp}$\cdot s_0$ or \texttt{ampstrict}. The number of full oscillations is specified by 
\texttt{kbperiod}.
If the variable \texttt{lconstvol} is T, the wall distance in the X and Y directions is also changed to maintain 
the constant volume of the box. 
After oscillations (or after equilibration if \texttt{loscillate} is F), the system is equilibrated again (with 
the resting time \texttt{ktrest}) and then the walls in the Z direction extend to $2s_0$ (if the variable 
\texttt{lpullfin} is T, otherwise the box does not deform anymore).

\subsection{Boundary conditions}

As mentioned before, by default the proteins are in an infinite space. Setting \texttt{lwall} T makes the 
simulation box finite. Then, by default the two walls perpendicular to the Z axis are ,,solid''. Setting \texttt{lpbc} 
T results in periodic boundary conditions only in the X and Y dimensions. To make them in all the three dimensions, 
the variable \texttt{lnowal} must be also set to T.

The variable \texttt{lwalls} is used to get ,,solid'' walls in the X and Y directions. Here ,,solid'' means repulsive, 
as decribed by the potential
$V_{wall}=\frac{-\epsilon}{9d^9}$, where $d$ is the distance from a residue to the wall \cite{virus1}. 

The walls in the $Z$ direction may have different ways of interacting with the residues. However, initially they 
always interact via $V_{wall}$ and there is a specific time when other special types of interaction may be turned 
on for the two walls along the Z axis. This specific time is controlled by the variable \texttt{kconnecttime}. 
Here are its possible values:
\begin{table}[h!]
    \begin{tabular}{|r|l|}
    \hline
    \texttt{kconnecttime} & description \\ \hline
    1 & just after reaching target density \texttt{tdens} \\ \hline
    3 & after waiting for \texttt{ktrest} after reaching \texttt{tdens} \\ \hline
    5 & just after reaching the maximal amplitude of oscillations \\ \hline
    7 & just after oscillations \\ \hline
    9 & never \\ \hline
    \end{tabular}
    \caption{Meaning of the kconnecttime variable}
\end{table}

Below are listed all the tested ways to introduce interactions
 between the residues and the two walls that perpendicular to the $Z$ direction.

\subsubsection{FCC walls}

If the variable \texttt{lfcc} is T, each of the walls is made from two layers of beads arranged in a 
face-centered-cubic lattice cut through the [111] crystallographic direction. The lattice constant is \texttt{af}. 
The beads are immovable (except for the movement of the whole wall) and interact via the L-J potential with the same 
amplitude as for disulfide bonds, \texttt{dislj}, and the same minimum, \texttt{walmindist}. The interactions are 
quasi-adiabatically turned on when the residue comes closer than \texttt{walmindist} (the minimum of the L-J potential) 
to one of the beads and is broken in the same way when the distance 
$r_{i,\text{bead}} > \texttt{cntfct} \cdot 2^{-1/6}\cdot \text{\texttt{walmindist}}$. 

\subsubsection{Flat attractive walls}

If \texttt{ljwal} is T, the residues interact with the wall via the same quasi-adiabatic L-J potential as above. 
The only difference is that when the residue comes closer than \texttt{walmindist} to the wall, it starts to be 
attracted by an artificial bead with the same X and Y coordinates as the residue's coordinates when it crosses the 
\texttt{walmindist} distance. This artificial bead does not attract other residues and disappears when 
$r_{i,\text{bead}} > \texttt{cntfct} \cdot 2^{-1/6}\cdot \text{\texttt{walmindist}}$. This can give a significant 
computational speed-up, as those artificial beads are not checked when updating the Verlet list.

If the variables \texttt{ljwal} and \texttt{lfcc} are F, all the residues closer than \texttt{walmindist} to the 
wall are harmonically attached to it, with the spring constant \texttt{HH1} and the equilibrium distance 
equal to the distance in the moment of turning on the interaction. The default value of \texttt{HH1} is 
30 $\epsilon/$\AA$^2$, which corresponds to strong, covalent bonding. 

\section{Advantages and disadvantages of the program}
The quasi-adiabatic and pseudo-improper-dihedral potentials are specifically designed for simulating intrinsically disordered proteins. The first potential generalizes the idea of the contact interactions taken from the structure-based Go model \cite{Sikora} (which is also included in the program). This enables the simulation of partially disordered proteins, like the structured domains connected by flexible linkers  (e.g. cellulosome proteins \cite{linkers}), proteins with disordered tracts (e.g. huntingtin \cite{tracts}) and large protein complexes with different amounts of complexity (e.g. gluten \cite{wieser}).
Our program was used to simulate the last case, gluten, which (according to our knowledge) was never simulated before \cite{gluten}. The program was also used to study aggregation of polyglutamine, which is a fully disordered protein \cite{polyq}. 
These two applications (both involving simulations of thousands of residues for more than a milisecond) prove the suitability of the quasi-adiabatic potential for large protein systems.
One disadvantage of the program is that (in the present form) it 
is not yet prepared to study folding-upon-binding mechanisms \cite{bind}.

The main strength of the program is allowing the study of large protein systems (in relatively long timescales), while retaining their protein nature. For the largest systems (like gluten) the faster quasi-adiabatic potential is best-suited \cite{lmc2018}. Smaller systems can be studied with better accuracy by the slower pseudo-improper-dihedral potential \cite{lmc2020}. The program may be also used to study fully structured proteins \cite{Hoang1,szymczak,Sikora,capsid,virwolek,hoangstiffness,wolek2,Zhao2,
Galera1,Galera2}

There are many other packages that may be used for implicit solvent, coarse-grained protein molecular dynamics \cite{gromacs,lammps,unres,openmm,cafemol,hoomd,mbn,desmond}, but none of them implements the quasi-adiabatic and pseudo-improper-dihedral potentials presented here. Many of them allow user-defined force fields \cite{gromacs,lammps,openmm,hoomd,desmond} and/or are open-source \cite{unres,openmm,cafemol}, however, none of them allows for an easy addition of dynamic contacts, switched on and off based on criteria of distance, direction and coordination number (see section \ref{quasi}). We discuss the advantages of using those criteria in a one-bead-per-residue model elsewhere \cite{lmc2018}. The novel pseudo-improper-dihedral potential (described in section \ref{pidmodel} and in \cite{lmc2020}) also cannot be easily added to the existing packages (with the possible exception of \cite{openmm} and \cite{cafemol}). The main disadvantage of using our software is that in 
the current state it is not as scalable as the alternatives (it cannot be effectively run on GPUs and it does not support MPI, only openMP).

\section*{Acknowledgments}
It is a pleasure to acknowledge interactions with A. Gomez-Sicilia,
T.~X. Hoang, M. Kogut, M. Raczkowski, M.~O. Robbins, B. R{\'{o}}{\.{z}}ycki, M. Sikora, P. Szymczak, 
J.~I. Su{\l}kowska, M. Wojciechowski, K. Wo{\l}ek and Y. Zhao
over two decades of developing the numerical approach presented here.
This research has received support from the National Science Centre (NCN),
Poland, under grant No. 2018/31/B/NZ1/00047 and the European H2020
FETOPEN-RIA-2019-01 grant PathoGelTrap No. 899616. The computer resources
were supported by the PL-GRID infrastructure.
\\

*Electronic address: l.mioduszewski@uksw.edu.pl

\end{document}